\documentclass[twocolumn,aps,prl,amsmath,amssymb]{revtex4-1}
\usepackage[latin9]{inputenc}
\usepackage{mathrsfs}
\usepackage{amsbsy}
\usepackage{amstext}
\usepackage{amssymb}
\usepackage{graphicx}
\usepackage{esint}

\makeatletter
\usepackage{epsfig}
\usepackage{bm}
\usepackage{dcolumn}
\usepackage{color}

\makeatother

\begin{document}

\title{Reduced quantum anomaly in a quasi-2D Fermi superfluid: The significance
of the confinement-induced effective range of interactions}

\author{Hui Hu$^{1}$, Brendan C. Mulkerin$^{1}$, Umberto Toniolo$^{1}$,
Lianyi He$^{2}$, and Xia-Ji Liu$^{1}$ }

\affiliation{$^{1}$Centre for Quantum and Optical Science, Swinburne University
of Technology, Melbourne, Victoria 3122, Australia}

\affiliation{$^{2}$Department of Physics and State Key Laboratory of Low-Dimensional
Quantum Physics, Tsinghua University, Beijing 100084, China}

\date{\today}
\begin{abstract}
A two-dimensional (2D) harmonically trapped interacting Fermi gas
is anticipated to exhibit a quantum anomaly and possesses a breathing
mode at frequencies different from a classical scale invariant value
$\omega_{B}=2\omega_{\perp}$, where $\omega_{\perp}$ is the trapping
frequency. The predicted maximum quantum anomaly ($\sim10\%$) has
not been confirmed in experiments. Here, we theoretically investigate
the zero-temperature density equation of state and the breathing mode
frequency of an interacting Fermi superfluid at the dimensional crossover
from three to two dimensions. We find that the simple model of a 2D
Fermi gas with a single $s$-wave scattering length is not adequate
to describe the experiments in the 2D limit, as commonly believed.
A more complete description of quasi-2D leads to a much weaker quantum
anomaly, consistent with the experimental observations. We clarify
that the reduced quantum anomaly is due to the significant confinement-induced
effective range of interactions, which is overlooked in previous theoretical
and experimental studies.
\end{abstract}


\maketitle
In strongly interacting quantum many-body systems, scale invariance
can lead to non-trivial consequences. An intriguing example is a three-dimensional
(3D) unitary Fermi gas with an infinitely large $s$-wave scattering
length $a_{3D}=\pm\infty$ \cite{Werner2006}. At zero energy, the
free space eigenstates of a unitary Fermi gas have a scale-invariant
form, i.e., under a rescaling of the spatial coordinates $\vec{X}\rightarrow\vec{X}/\lambda$,
the scaled wave functions satisfy $\psi(\vec{X}/\lambda)=\lambda^{-\nu}\psi(\vec{X})$
for any scaling factor $\lambda>0$. In the presence of an isotropic
harmonic trap of frequency $\omega_{0}$, a set of trap eigenstates
can then be constructed from zero-energy states in free space \cite{Werner2006},
whose spectrum form a ladder with a step of $2\hbar\omega_{0}$, indicating
the existence of a well-defined quasiparticle (i.e., breathing mode)
even in the strongly correlated regime. This non-trivial exact mode
can be understood from a hidden $SO(2,1)$ symmetry in the problem
\cite{Pitaevskii1997}.

\begin{figure}
\centering{}\includegraphics[width=0.48\textwidth]{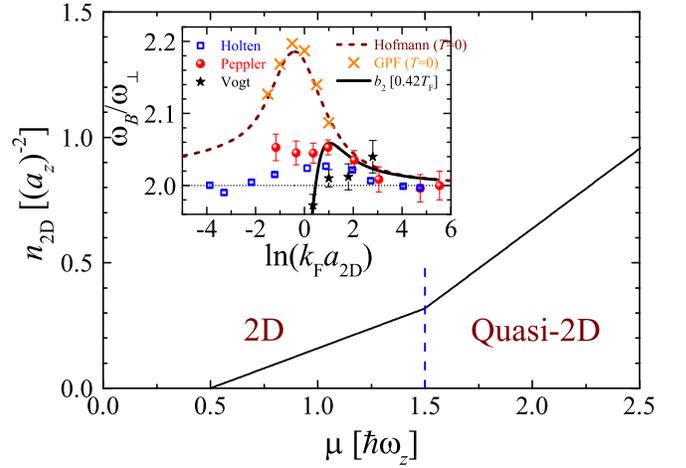}\caption{\label{fig1_idealgas} The column density $n_{2D}=\int dzn(z)$ of
an ideal Fermi gas, in units of $a_{z}^{-2}=M\omega_{z}/\hbar$, at
the dimensional crossover from 2D to 3D. The inset shows the predicted
breathing mode frequencies of a 2D interacting Fermi gas at $T=0$
using QMC EoS (dashed line) \cite{Hofmann2012} and GPF EoS (crosses)
calculated in this work, and at $T=0.42T_{F}$ using virial expansion
(solid line) \cite{Mulkerin2018}, compared with the experimental
data by Vogt \textit{et al.} (stars, $0.42T_{F}$) \cite{Vogt2012},
Holten \textit{et al.} (squares, $0.10-0.18T_{F}$) \cite{Holten2018}
and Peppler \textit{et al.} (circles, $0.14-0.22T_{F}$) \cite{Peppler2018}.}
\end{figure}

Classically, a two-dimensional (2D) atomic gas interacting through
a contact interaction is also scale invariant. The hidden $SO(2,1)$
symmetry under an isotropic trap (of frequency $\omega_{\bot}$) would
similarly lead to an exact breathing mode with frequency $\omega_{B}=2\omega_{\perp}$,
for both bosons and fermions \cite{Pitaevskii1997}. Quantum mechanically,
however, the contact interaction needs renormalization and the bare
interaction strength should be replaced by a regularized 2D $s$-wave
scattering length $a_{2D}$ \cite{Adhikari1986}. As a result of this
new length scale, scale invariance of 2D quantum gases explicitly
breaks down \cite{Holstein1993} and the breathing mode frequency
should depend on $a_{2D}$. In a 2D weakly interacting Bose gas, the
quantum anomaly is too weak to be observed \cite{Olshanii2010,Hu2011,Merloti2013}.
For an interacting 2D Fermi gas, the predicted quantum anomaly, i.e.,
$\delta\omega_{B}/(2\omega_{\bot})$, is significant and can reach
approximately $10\%$ in the strongly interacting crossover regime
at zero temperature \cite{Hofmann2012,Taylor2012,Cao2012}, as shown
in the inset of Fig. \ref{fig1_idealgas} as a function of $\ln(k_{F}a_{2D})$,
where $k_{F}$ is the Fermi wavevector at the trap center. The 2D
regime can be experimentally realized by imposing a tight axial confinement
with a large trap aspect ratio $\lambda=\omega_{z}/\omega_{\bot}$
\cite{Turlapov2017,Vogt2012,Makhalov2014,Holten2018,Peppler2018},
when the number of atoms $N$ is sufficiently small and only the ground
single-particle state in the axial direction is populated \cite{Turlapov2017}.
For an ideal Fermi gas, this requires $N<N_{2D}=\lambda^{2}$ or equivalently
a chemical potential $\mu<1.5\hbar\omega_{z}$ (see Fig. 1) \cite{Turlapov2017}.

The prediction of the $10\%$ quantum anomaly, unfortunately, has
never been confirmed experimentally. The first experiment measured
an anomaly of less than $1\%$ at temperature $0.42T_{F}$ \cite{Vogt2012},
where $T_{F}$ is the Fermi temperature. While the discrepancy may
be understood as a temperature effect \cite{Chan2013,Mulkerin2018},
two most recent measurements \cite{Holten2018,Peppler2018} reported
consistently quantum anomaly of about $1.3\%$ and $2.5\%$, respectively,
at temperature as low as $\sim0.1T_{F}$ (see the inset of Fig. \ref{fig1_idealgas}).
The large discrepancy of the measurements compared to the predicted
anomaly is rather surprising. The purpose of this Letter is to show
that the puzzle can be resolved by including all the trapped single-particle
states along the axial direction and hence taking into account the
quasi-2D nature of the experimental setup, which leads to an \emph{unexpected}
large confinement-induced effective range of interactions $R_{s}$
in the 2D limit, a fact that is severely overlooked in the past studies
of a 2D strongly interacting Fermi gas.

\begin{figure}[t]
\begin{centering}
\includegraphics[width=0.45\textwidth]{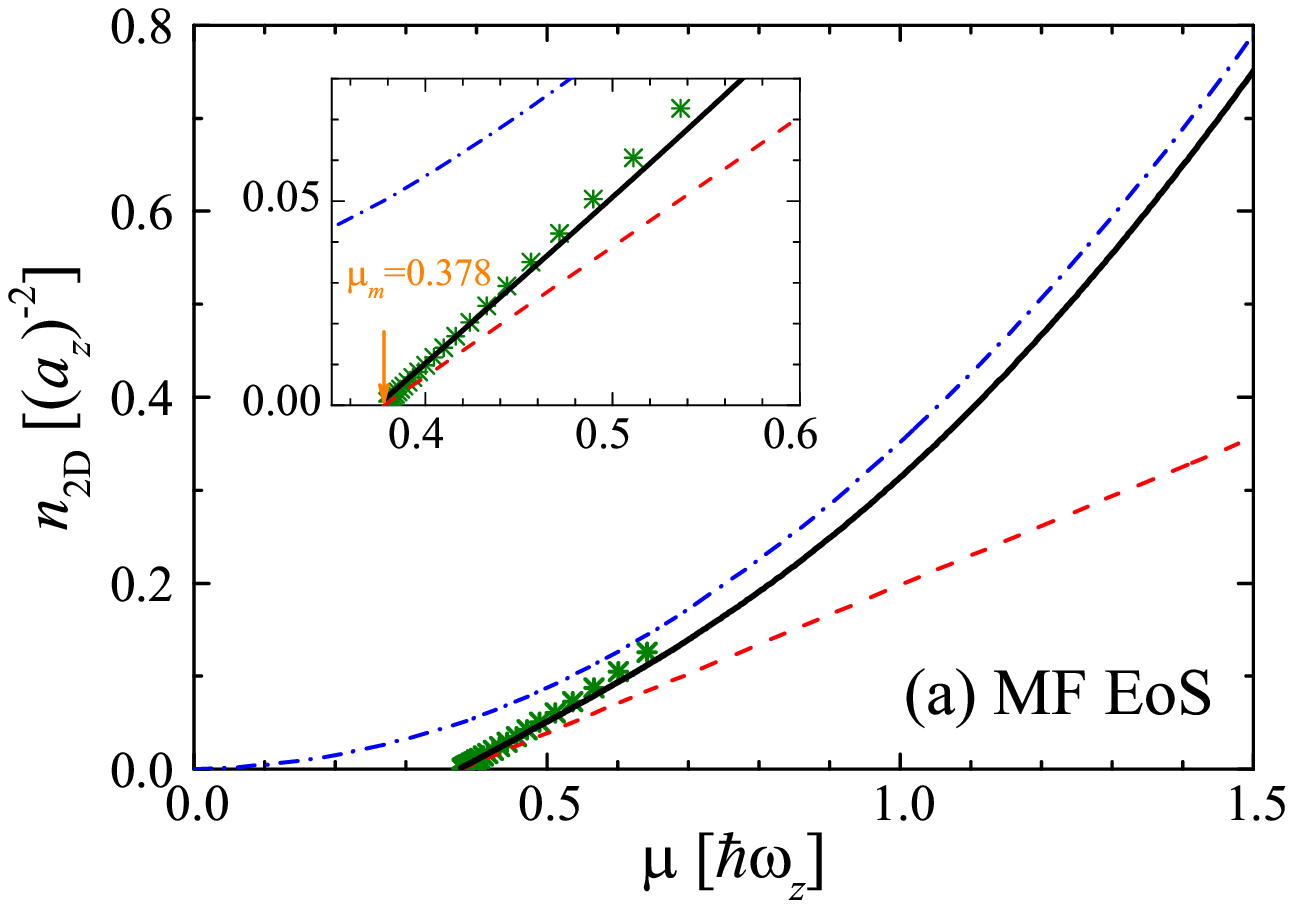} 
\par\end{centering}
\begin{centering}
\includegraphics[width=0.45\textwidth]{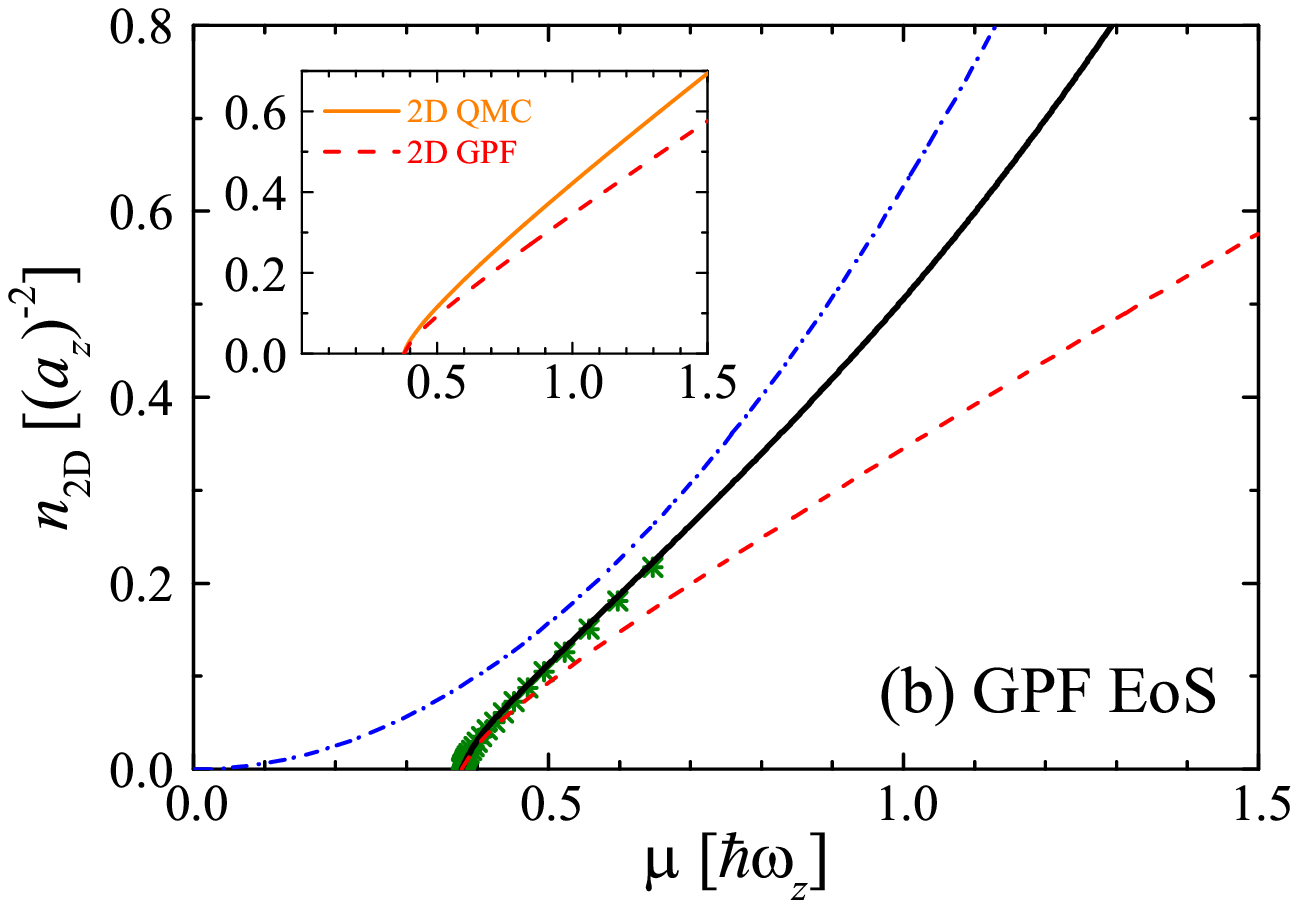} 
\par\end{centering}
\caption{\label{fig2_n2d_unitarygas} The density EoS of a unitary Fermi gas
at the dimensional crossover, calculated using the mean-field theory
(a) and GPF theory (b). The anticipated behavior in the 3D limit is
shown by the blue dot-dashed lines, and the predictions of the 2D
models without and with the effective range of interactions are plotted
by the red dashed lines and green asterisks, respectively. The inset
in (a) highlights the EoS near the minimum chemical potential $\mu_{m}=(\hbar\omega_{z}-\epsilon_{B})/2\simeq0.378\hbar\omega_{z}$.
The inset in (b) compares the 2D EoS with contact interactions, predicted
by QMC and the GPF theory.}
\end{figure}

\begin{figure}[t]
\centering{}\includegraphics[width=0.48\textwidth]{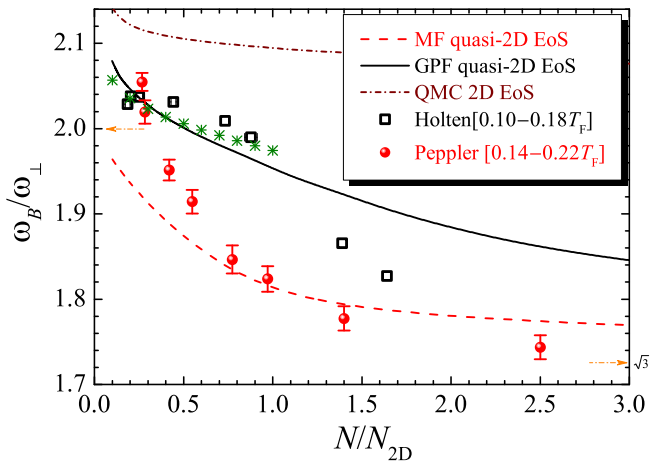}\caption{\label{fig3_wb_unitarygas} The breathing mode frequency of a unitary
Fermi gas at the dimensional crossover, as a function of $N/N_{2D}$.
The squares and circles are the experimental data, measured by Holten
\textit{et al.} \cite{Holten2018} at $a_{z}/a_{3D}\simeq-0.35$ and
Peppler \textit{et al.} \cite{Peppler2018} $a_{z}/a_{3D}=0$. The
GPF prediction of the two-channel 2D model with the effective range
of interactions is also shown by green asterisks (see text for details).}
\end{figure}

\begin{figure}[t]
\centering{}\includegraphics[width=0.48\textwidth]{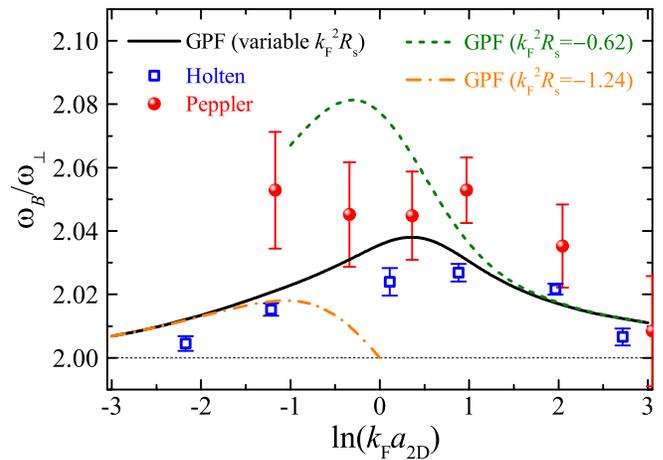}\caption{\label{fig4_wb_GPF2c} The breathing mode frequency of a strongly
interacting 2D Fermi gas at the lowest experimental number of atoms
$N/N_{2D}\sim0.2$. The squares and circles are the experimental data
by Holten \textit{et al.} ($0.10-0.18T_{F}$) \cite{Holten2018} and
Peppler \textit{et al.} (,$0.14-0.22T_{F}$) \cite{Peppler2018},
respectively. The lines show the predictions of the two-channel 2D
model at different effective range of interactions, within the GPF
theory.}
\end{figure}

Theoretically, the understanding of a strongly interacting Fermi gas
at the dimensional crossover is a highly non-trivial challenge, even
at the mean-field level, due to both infrared and ultraviolet divergences
at low and high energies, respectively \cite{Martikainen2005,Fischer2013}.
In this work, we completely solve the zero-temperature dimensional
crossover problem. In particular, we take into account strong Gaussian
pair fluctuations (GPF) on top of the mean-field solutions and therefore
\emph{quantitatively} determine the equation of state (EoS) and the
breathing mode of a strongly interacting Fermi gas at the 2D-3D crossover
(see Figs. \ref{fig2_n2d_unitarygas} and \ref{fig3_wb_unitarygas}).
We find surprisingly that, in sharp contrast to the common belief,
the Fermi cloud in the 2D limit can not be adequately described by
the simple 2D model with a single scattering length $a_{2D}$. At
the lowest experimental number of atoms $N/N_{2D}\sim0.2$, the dimensionless
effective range of interactions $k_{F}^{2}R_{s}\sim-\sqrt{N/N_{2D}}=O(1)$
is comparable in magnitude to the interaction parameter $\ln(k_{F}a_{2D})$
in the strongly interacting regime, leading to a much reduced quantum
anomaly as experimentally observed (see Fig. \ref{fig4_wb_GPF2c}).

\textit{Theoretical framework}. \textemdash{} The experimentally realized
quasi-2D Fermi gas of $^{6}$Li or $^{40}$K atoms near a broad Feshbach
resonance \cite{Vogt2012,Holten2018,Peppler2018} can be described
by \cite{Liu2005}, 
\begin{equation}
\mathcal{H}={\displaystyle {\displaystyle \sum_{\sigma}}}\psi_{\sigma}^{\dagger}(\mathbf{r})\mathcal{H}_{0}\psi_{\sigma}(\mathbf{r})+U\psi_{\uparrow}^{\dagger}(\mathbf{r})\psi_{\downarrow}^{\dagger}(\mathbf{r})\psi_{\downarrow}(\mathbf{r})\psi_{\uparrow}(\mathbf{r}),\label{eq:hami}
\end{equation}
where $\psi_{\sigma}(\mathbf{r})$ is the annihilation operator for
the spin state $\sigma=\uparrow,\downarrow$ at position $\mathbf{r}=(\mathbf{\boldsymbol{\rho}},z)$,
$\mathcal{H}_{0}=-\hbar^{2}\nabla^{2}/(2M)+M(\omega_{\perp}^{2}\boldsymbol{\rho}^{2}+\omega_{z}^{2}z^{2})/2-\mu_{\textrm{g}}$
is the single-particle Hamiltonian with atomic mass $M$, $\mu_{\textrm{g}}$
is the chemical potential, and $U$ denotes the contact interaction
strength and should be regularized by $a_{3D}$ via, $M/(4\pi\hbar^{2}a_{3D})=1/U+\sum_{\mathbf{k}}M/(\hbar^{2}\mathbf{k}^{2})$.
As the transverse trapping potential $M\omega_{\perp}^{2}\boldsymbol{\rho}^{2}/2$
varies slowly in real space, it is convenient to use the local density
approximation (LDA) and define a local chemical potential $\mu(\mathbf{\boldsymbol{\rho}})=\mu_{\textrm{g}}-M\omega_{\perp}^{2}\boldsymbol{\rho}^{2}/2$
\cite{Butts1997}. In the following, we first treat a locally transversely
homogeneous Fermi gas with $\mu(\mathbf{\boldsymbol{\rho}})$, in
which the single-particle wave-function takes a plane-wave form $\propto\exp(i\mathbf{k}\cdot\boldsymbol{\rho})$
with wave vector $\mathbf{k}$ in the transverse direction.

At the mean-field level, to account for the tight axial confinement,
we solve the inhomogeneous Bogoliubov-de Gennes (BdG) equation \cite{Liu2007a,Liu2007b},
\begin{equation}
\left[\begin{array}{cc}
\mathcal{H}_{0}(\mathbf{k}) & \Delta(z)\\
\Delta^{*}(z) & -\mathcal{H}_{0}(\mathbf{k})
\end{array}\right]\left[\begin{array}{c}
u_{\eta\mathbf{k}}(z)\\
v_{\eta\mathbf{k}}(z)
\end{array}\right]=E_{\eta\mathbf{k}}\left[\begin{array}{c}
u_{\eta\mathbf{k}}(z)\\
v_{\eta\mathbf{k}}(z)
\end{array}\right],\label{eq:BdG}
\end{equation}
for the quasiparticle wave functions $u_{\eta\mathbf{k}}(z)\exp(i\mathbf{k}\cdot\boldsymbol{\rho})$
and $v_{\eta\mathbf{k}}(z)\exp(i\mathbf{k}\cdot\boldsymbol{\rho})$
with energy $E_{\eta\mathbf{k}}>0$. Here, $\mathcal{H}_{0}(\mathbf{k})\equiv-[\hbar^{2}/(2M)]d^{2}/dz^{2}+\hbar^{2}\mathbf{k}^{2}/(2M)-\mu(\boldsymbol{\rho})$
and we have used $\eta$ to explicitly index the energy spectrum for
a given wave vector $\mathbf{k}$. The pairing field $\Delta(z)$
in the BdG equation should be determined self-consistently, according
to $\Delta(z)=U\sum_{\eta\mathbf{k}}u_{\eta\mathbf{k}}(z)v_{\eta\mathbf{k}}^{*}(z)$.
The resulting mean-field column density is then given by, $n_{2D}^{(\textrm{MF})}=\int dzn_{\textrm{MF}}(z)=2\int dz\sum_{\eta\mathbf{k}}v_{\eta\mathbf{k}}(z)v_{\eta\mathbf{k}}^{*}(z)$
\cite{SUPP}.

In the strongly interacting regime, mean-field theory is qualitatively
reliable only. For a quantitative description, we must go beyond mean
field and include strong pair fluctuations by generalizing the GPF
theory \cite{Hu2006,Hu2007,Diener2008,He2015,Toniolo2017,Mulkerin2017}
to the case of an \emph{inhomogeneous} pairing field. This non-trivial
generalization is achieved by working out the vertex function $\Gamma(\mathbf{q},$$i\nu_{l})$
(i.e., the Green function of Cooper pairs) and the associated thermodynamic
potential $\Omega_{\textrm{GF}}=(k_{B}T/2)\sum_{\mathcal{\mathscr{Q}}\equiv(\mathbf{q},i\nu_{l})}\ln[-\Gamma^{-1}(\mathcal{\mathscr{Q}})]$,
where $\nu_{l}=2\pi lk_{B}T$ is the bosonic Matsubara frequency.
In greater detail, we have $\Omega_{\textrm{GF}}=k_{B}T\sum_{\mathcal{\mathscr{Q}}}\mathcal{S}(\mathcal{\mathscr{Q}})e^{i\nu_{l}0^{+}}$
\cite{Diener2008,Mulkerin2017}, 
\[
\mathcal{S}\left(\mathcal{\mathscr{Q}}\right)=\frac{1}{2}\ln\left[1-\frac{M_{12}^{2}\left(\mathcal{\mathscr{Q}}\right)}{M_{11}\left(\mathcal{\mathscr{Q}}\right)M_{11}\left(-\mathcal{\mathscr{Q}}\right)}\right]+\ln M_{11}\left(\mathcal{\mathscr{Q}}\right),
\]
and the matrix elements $M_{11}(\mathcal{\mathscr{Q}})$ and $M_{12}(\mathcal{\mathscr{Q}}$)
of the inverse vertex function $\Gamma^{-1}(\mathcal{\mathscr{Q}})$
can be written in terms of the inhomogeneous BCS Green function of
fermions \cite{SUPP}. Once we obtain $\Omega_{\textrm{GF}}$, we
calculate the column density $n_{2D}^{(\textrm{GF})}=-\partial\Omega_{\textrm{GF}}/\partial\mu(\boldsymbol{\rho})$.

\textit{Universal EoS at the dimensional crossover}. \textemdash{}
Using the mean-field theory or GPF theory, we calculate the column
density $n_{2D}=n_{2D}^{(\textrm{MF})}$ or $n_{2D}=n_{2D}^{(\textrm{MF})}+n_{2D}^{(\textrm{GF})}$
at a given local chemical potential $\mu\equiv\mu(\boldsymbol{\rho})$.
Focusing on the \emph{unitary} limit where $a_{3D}\rightarrow\pm\infty$,
the zero-temperature results are shown in Fig. \ref{fig2_n2d_unitarygas}
by black solid lines. This unitary limit is of particular interest,
as the length scale $a_{3D}$ in the interatomic interaction disappears
and the system therefore should exhibit universal thermodynamics \cite{Hu2007,Ho2004,Nascimbene2010,Navon2010,Ku2012}.
In our case, we can express $n_{2D}$ as a function of $\mu/(\hbar\omega_{z})$
only and the predicted universal EoS in Fig. \ref{fig2_n2d_unitarygas}
could be experimentally determined by a single-shot measurement of
the column density at the lowest attainable temperature \cite{Navon2010}.

In the 3D limit, where a number of singe-particle levels in the axial
direction are occupied, we may use the LDA to handle the axial trap
$M\omega_{z}^{2}z^{2}/2$. This gives rise to \cite{SUPP} 
\begin{equation}
n_{2D}\left(\mu\gg\hbar\omega_{z}\right)=\frac{1}{2\pi^{2}\xi^{3/2}}\left(\frac{M\mu^{2}}{\hbar^{3}\omega_{z}}\right),\label{eq:n2D_unitary_3D}
\end{equation}
where $\xi$ is the so-called Bertsch parameter. The mean-field and
GPF theories predict $\xi_{BCS}\simeq0.59$ and $\xi_{GPF}\simeq0.40$,
respectively. The latter is very close to the latest experimental
value $\xi_{exp}=0.376(5)$ \cite{Ku2012}. In the opposite 2D limit,
if we use a simple 2D model with contact interactions \cite{He2015,Bertaina2011},
the mean-field theory provides a simple EoS, $n_{2D}(\mu\rightarrow\mu_{m})=M(\mu-\mu_{m})/(\pi\hbar^{2})$
\cite{He2015}, where $\mu_{m}=(\hbar\omega_{z}-\epsilon_{B})/2$
is the minimum chemical potential allowed, due to the existence of
a two-body bound state with binding energy $\epsilon_{B}=\hbar^{2}/(Ma_{2D}^{2})$
\cite{SUPP,Petrov2001}. More accurate EoS in the 2D limit could be
obtained using numerically exact quantum Monte Carlo (QMC) simulations
\cite{Bertaina2011,Shi2015} or the approximate GPF theory \cite{He2015},
as illustrated in the inset of Fig. \ref{fig2_n2d_unitarygas}(b).
The relative difference between QMC and GPF results is small (i.e.,
less than $15\%$), suggesting that the GPF theory is quantitatively
reliable also in the 2D limit \cite{NoteGPF}.

In Fig. \ref{fig2_n2d_unitarygas}, we show the anticipated EoSs in
the 3D and 2D limits with contact interactions using blue dot-dashed
lines and red dashed lines, respectively. Our predicted EoS at the
dimensional crossover (black curves), from both mean-field and GPF
theories, lies in between and seems to smoothly connect the two limits.
However, a close examination of the 2D limit shows that the anticipated
2D EoS with a single $s$-wave scattering length $a_{2D}$ cannot
fully account for the predicted quasi-2D results.

This is clearly seen from the mean-field EoS. In the inset of Fig.
\ref{fig2_n2d_unitarygas}(a), we highlight the density EoS near the
2D limit. Although the predicted mean-field EoS $n_{2D}^{(\textrm{MF})}$
shows the expected linear dependence on $M(\mu-\mu_{m})/\hbar^{2}$,
the slope of the curve is significantly larger than $1/\pi$ from
the simple 2D model of contact interactions. Therefore, it is evident
that the 2D model with a single parameter $a_{2D}$ fails to adequately
describe the EoS near the 2D limit. A hint for this failure actually
was already observed in the first measurement of the ground state
EoS of an interacting 2D Fermi gas \cite{Makhalov2014}, where the
definition of $a_{2D}$ should be modified to explain the discrepancy
between the experimental data and the QMC prediction \cite{SUPP}. 

A new effective 2D model Hamiltonian therefore has to be introduced,
with additional terms accounting for the enhanced slope in the quasi-2D
EoS in the strongly interacting regime. As a minimum setup, we consider
the inclusion of the effective range of interactions induced by the
tight harmonic confinement. Indeed, by expanding the expression of
the quasi-2D scattering amplitude $f_{Q2D}(k)$ first calculated by
Petrov and Shlyapnikov \cite{Petrov2001} to the order $O(k^{2})$
\cite{SUPP},
\begin{equation}
f_{Q2D}\left(k\right)=-\frac{2\pi}{\ln\left(ka_{2D}\right)+R_{s}k^{2}/2-i\pi/2+\cdots},
\end{equation}
we find an effective range of interactions, $R_{s}=-a_{z}^{2}\ln2$,
which is ignored in most of previous studies. This is a surprisingly
large effective range, if we consider the typical Fermi wavevector
$k_{F}\sim a_{z}^{-1}$ and scattering length $a_{2D}\sim a_{z}$,
and hence $\ln(k_{F}a_{2D})\sim R_{s}k_{F}^{2}$ \cite{SUPP}. More
precisely, by taking a peak density of an \emph{ideal} trapped 2D
Fermi gas $n_{2D}=(\sqrt{N}/\pi)(M\omega_{\perp}/\hbar)=a_{z}^{-2}\sqrt{N/N_{2D}}/\pi=k_{F}^{2}/(2\pi)$
\cite{Turlapov2017}, we obtain a dimensionless effective range $k_{F}^{2}R_{s}=-(2\ln2)\sqrt{N/N_{2D}}=O(1)$
at the realistic experimental number of atoms $N\gtrsim0.2N_{2D}$
\cite{Holten2018,Peppler2018}.

We have developed a two-channel 2D model to account for the effective
range of interactions (see Supplemental Material \cite{SUPP} for
details and also Refs. \cite{Kestner2007} and \cite{Zhang2008}).
The resulting mean-field and GPF predictions for the density EoS are
shown in Fig. \ref{fig2_n2d_unitarygas} by green asterisks. In the
2D limit (i.e., $\mu\rightarrow\mu_{m}$), we find excellent agreement
between the full quasi-2D simulations and the two-channel calculations,
confirming the importance of the effective range. As we shall see,
it is also responsible for the much reduced quantum anomaly in the
breathing mode frequency.

\textit{Breathing mode frequency}. \textemdash{} In the strongly interacting
regime, the breathing mode can be well-described by a hydrodynamic
theory \cite{Taylor2008}, which has been successfully applied to
predict a large variety of collective oscillations in both Fermi and
Bose gases \cite{Menotti2002,Hu2004,Taylor2009}. Here, it is convenient
to use the well-documented sum-rule approach \cite{Menotti2002},
which leads to 
\begin{equation}
\hbar^{2}\omega_{B}^{2}=-2\left\langle \rho^{2}\right\rangle \left[\frac{d\left\langle \rho^{2}\right\rangle }{d\left(\omega_{\perp}^{2}\right)}\right]^{-1},
\end{equation}
where $\left\langle \rho^{2}\right\rangle =N^{-1}\int d^{2}\boldsymbol{\rho}[\rho^{2}n_{2D}(\boldsymbol{\rho})]$
is the squared radius of the Fermi cloud and the chemical potential
$\mu_{\textrm{g}}$ in the local chemical potential $\mu(\boldsymbol{\rho})$
should be adjusted to satisfy the number equation, $N=\int d^{2}\boldsymbol{\rho}n_{2D}(\boldsymbol{\rho})$.
We note that, the breathing mode frequency evaluated using the sum-rule
approach is \emph{exact} when the density EoS takes a polytropic form,
i.e., $\mu(n_{2D})\propto(n_{2D})^{\gamma}$. In that case, the density
profile is easy to determine within LDA and one finds $\omega_{B}/\omega_{\perp}=\sqrt{2+2\gamma}$
\cite{Menotti2002}.

For a quasi-2D unitary Fermi gas in the 3D limit, the density EoS
is precisely described by a polytropic form with $\gamma=1/2$, as
given in Eq. (\ref{eq:n2D_unitary_3D}), and we obtain $\omega_{B}=\sqrt{3}\omega_{\perp}$
\cite{Hu2014,DeRosi2015}. On the contrary, in the 2D limit the mean
field theory predicts a classical EoS $\mu(n_{2D})-\mu_{m}=$$\pi\hbar^{2}n_{2D}/M$
with $\gamma=1$, and hence we recover the scale-invariant result
$\omega_{B}=2\omega_{\perp}$. Quantum fluctuations upshift the breathing
mode frequency and lead to the quantum anomaly \cite{Hofmann2012,Taylor2012}.

At the dimensional crossover, we report in Fig. \ref{fig3_wb_unitarygas}
the breathing mode frequency of a unitary Fermi gas as a function
of $N/N_{2D}$, calculated using both the quasi-2D mean-field (red
dashed line) and GPF theories (black solid line), and compare them
with the recent measurements at Heidelberg \cite{Holten2018} and
at Swinburne \cite{Peppler2018}. We also show the result obtained
by using the QMC EoS of the simple 2D model of contact interactions
\cite{Shi2015} (brown dot-dashed line) and the GPF prediction of
the two-channel 2D model (green asterisks). The mode frequencies found
by our quasi-2D and two-channel 2D calculations, and measured by experiments
all exhibit a strong dependence on $N/N_{2D}$, in sharp contrast
to the pure 2D QMC prediction. In particular, the anticipated 2D behavior,
i.e., the $\sim10\%$ upshift of the mode frequency, is already washed
out at a small number of atoms $N/N_{2D}\sim0.2$, due to the significant
effective range of interactions. As the number of atoms increase,
the data from the Heidelberg group \cite{Holten2018} follows continuously
our GPF prediction; however, the measurement at Swinburne \cite{Peppler2018}
agrees better with the mean-field result. The source for such a difference
requires a further study.

To confirm conclusively that the observed reduced quantum anomaly
arises from the large effective range, we compare in Fig. \ref{fig4_wb_GPF2c}
the GPF prediction of the two-channel 2D model with the experimental
data at $N/N_{2D}\sim0.2$, as a function of the interaction parameter
$\ln(k_{F}a_{2D})$. On the weak coupling side (i.e., $\ln(k_{F}a_{2D})>1$),
the result with $k_{F}^{2}R_{s}=-(2\ln2)\sqrt{N/N_{2D}}\simeq-0.62$
(dash green line) agrees with the data \cite{Holten2018,Peppler2018}.
However, towards the strong coupling regime, our prediction over-estimates
the shift. This is easy to understand, since in that limit the peak
density of the Fermi cloud should be much larger than the peak density
of an ideal 2D Fermi gas that we have assumed. To account for this
effect, we may assume that the peak density doubles for strong-coupling
\cite{Martiyanov2016} and take an interpolative dimensionless effective
range \cite{SUPP}, $k_{F}^{2}R_{s}=-0.62(1+1/[(k_{F}a_{2D})^{2}+1])$.
The resulting frequency (black line) fits reasonably well with the
experimental data at all interaction strengths. 

\textit{Conclusions}. \textemdash{} We have developed a strong-coupling
theory for an interacting Fermi gas at the dimensional crossover from
3D to 2D. We have clarified that in the 2D limit under the current
experimental conditions, a confinement-induced effective range of
interactions is very significant and should be accounted for both
theoretically (i.e., via a two-channel 2D model) and experimentally.
It leads to a much reduced quantum anomaly, as observed in the two
most recent measurements \cite{Holten2018,Peppler2018}. The consequence
of such a large effective range in other quantum phenomena, for example,
the Berezinskii-Kosterlitz-Thouless transition \cite{Mulkerin2017,Berezinskii1972,Kosterlitz1973},
remains to be understood.
\begin{acknowledgments}
We thank Paul Dyke for useful discussions and for sharing the experimental
data. Our research was supported by Australian Research Council's
(ARC) Programs FT130100815 and DP170104008 (HH), FT140100003 and DP180102018
(XJL), the National Natural Science Foundation of China, Grant No.
11775123 (LH), and the National Key Research and Development Program
of China, Grant No. 2018YFA0306503 (LH). 
\end{acknowledgments}


\begin{thebibliography}{10}
\bibitem{NoteGPF}For the breathing mode frequency of an interacting
2D Fermi gas, the predictions obtained by using QMC EoS and GPF EoS
are compared in the inset of Fig. \ref{fig1_idealgas}. It is readily
seen that the GPF theory provides a quantitative description of the
2D breathing mode frequency in the strongly interacting crossover
regime.
\end{thebibliography}
\end{document}